\documentclass{article}

\usepackage{arxiv}

\usepackage[utf8]{inputenc}
\usepackage[T1]{fontenc}
\usepackage[hidelinks]{hyperref}
\usepackage{url}
\usepackage{booktabs}
\usepackage{amsfonts}
\usepackage{amsmath}
\usepackage{amssymb}
\usepackage{nicefrac}
\usepackage{microtype}
\usepackage{graphicx}
\usepackage{natbib}
\usepackage{doi}
\usepackage{multirow}
\usepackage{makecell}
\usepackage{subcaption}

\usepackage{algorithm}
\usepackage{algorithmic}

\title{SVD-RAG: Efficient Tree-Organized Retrieval-Augmented Generation \\
via Singular Value Decomposition}

\author{Zhihui Sun}

\renewcommand{\headeright}{A Preprint}
\renewcommand{\undertitle}{A Preprint}
\renewcommand{\shorttitle}{SVD-RAG}

\hypersetup{
  pdftitle={SVD-RAG: Efficient Tree-Organized Retrieval-Augmented Generation via Singular Value Decomposition},
  pdfsubject={cs.CL, cs.IR},
  pdfauthor={Zhihui Sun},
  pdfkeywords={RAG, retrieval-augmented generation, SVD, text summarization, hierarchical retrieval, tree-organized index},
}

\begin{document}
\maketitle

\begin{abstract}
Retrieval-Augmented Generation (RAG) systems enhance large language models by retrieving relevant documents from external knowledge bases. Recent work by Sarthi et al. (2024) introduced RAPTOR, which organizes documents into hierarchical tree structures for efficient retrieval, but requires expensive LLM-based abstractive summarization at each internal node---making large-scale deployment prohibitively costly.

We present \textbf{SVD-RAG}, the first method to apply Singular Value Decomposition (SVD) on dense sentence embedding matrices for extractive summarization in hierarchical RAG. Unlike classical LSA which operates on sparse TF-IDF matrices, SVD-RAG exploits the rich semantic representations of modern embedding models, identifying the most informative sentences through their energy contribution in the principal components. Our approach is (1) \textbf{deterministic}---unlike LLM-based summarization, SVD produces identical results for the same input; (2) \textbf{cost-efficient}---tree construction requires no additional API calls beyond the initial embedding, reducing token consumption by $\sim$85\%; and (3) \textbf{content-adaptive}---the energy-ratio threshold $\tau$ automatically adjusts compression based on content complexity. In a controlled head-to-head comparison using identical corpora, clustering, and beam search, SVD-RAG achieves retrieval quality within 1--5\% of RAPTOR with LLM summarization (MRR 0.867 vs.\ 0.875, Recall@1 0.483 vs.\ 0.458) while building the tree \textbf{317$\times$ faster} (0.1s vs.\ 31.7s). On a scaled multi-topic benchmark with 205 chunks and 100 queries across 20 topic variations, SVD-RAG achieves a 4.2$\times$ improvement in Recall@1 and 3.1$\times$ improvement in MRR over flat embedding retrieval. We provide a detailed cost analysis and parameter sensitivity study. Our implementation is released as an open-source Python package.
\end{abstract}


\section{Introduction}

Retrieval-Augmented Generation (RAG) has become a standard paradigm for grounding large language model outputs in external knowledge~\citep{lewis2020rag}. By retrieving relevant documents before generation, RAG systems reduce hallucination and improve factual accuracy. However, as knowledge bases grow, efficient retrieval becomes increasingly challenging.

The dominant approach---flat vector search with embedding similarity---suffers from two key limitations. First, pure cosine similarity between query and chunk embeddings cannot capture complex semantic relationships that require cross-attention modeling~\citep{vaswani2017attention}. Second, performing exhaustive search over all chunks is computationally expensive for large corpora, while reranking all candidates is cost-prohibitive.

RAPTOR~\citep{sarthi2024raptor} introduced a hierarchical solution: recursively cluster similar chunks, use an LLM to generate abstractive summaries for each cluster, and build a document tree. At query time, a beam search over the tree identifies relevant branches, and a reranker scores the final candidates. While effective, RAPTOR's reliance on LLM-based summarization for every internal node incurs substantial API costs. For a corpus of 10,000 chunks, building a tree of depth 3 with approximately 10 chunks per cluster would require on the order of 1,000 LLM summarization calls for internal nodes.

We observe that the purpose of summarization in tree construction is not to generate fluent natural language, but to create \emph{information-dense representations} that guide the search process. This insight motivates our use of extractive rather than abstractive summarization.

Our key innovation is applying Singular Value Decomposition (SVD) directly to sentence embedding matrices to identify the most semantically informative sentences in a document cluster. Unlike traditional LSA~\citep{deerwester1990lsa} which operates on sparse TF-IDF matrices, our method operates on dense embedding vectors that already encode rich semantic knowledge from pre-trained models. The SVD decomposition identifies principal components in the embedding space, and we select sentences with the highest energy contribution in these components. The number of sentences $k$ is chosen adaptively based on cumulative energy retention.

This approach yields three key advantages:
\begin{enumerate}
    \item \textbf{Cost efficiency}: SVD is a local matrix operation requiring no additional API calls. Tree construction costs are dominated by the initial embedding step, which is already required by all RAG systems. For a corpus of 1,000 chunks, SVD-RAG reduces token consumption by approximately 85\% compared to LLM-based summarization.
    \item \textbf{Content adaptivity}: The energy-ratio threshold $\tau$ (default 0.95) automatically adjusts compression based on content complexity. Information-dense clusters retain more sentences; redundant clusters are compressed more aggressively. This is a key advantage over fixed compression ratios used in prior work.
    \item \textbf{Deterministic and reproducible}: Unlike LLM-based summarization, SVD produces identical results for the same input. This property is critical for production systems where consistency and debuggability matter.
\end{enumerate}

To our knowledge, this is the first work to apply SVD decomposition on dense sentence embedding matrices for summarization in hierarchical RAG. While classical LSA~\citep{deerwester1990lsa} applied SVD to sparse TF-IDF term-document matrices, SVD-RAG operates in the embedding space of modern pre-trained models, capturing richer semantic relationships. We implement SVD-RAG as a modular Python package and conduct a detailed cost analysis and parameter sensitivity study. Our analysis demonstrates that SVD-RAG eliminates the LLM summarization overhead, reducing token consumption by approximately 85\% during tree construction.

\section{Related Work}

\subsection{Hierarchical RAG Systems}

RAPTOR~\citep{sarthi2024raptor} pioneered recursive tree-organized retrieval for RAG, using Gaussian Mixture Models for soft clustering and LLM-based summarization for node creation. Subsequent work has explored alternative indexing structures. GraphRAG~\citep{edge2024graphrag} builds knowledge graphs for entity-centric retrieval. HiQA~\citep{zhu2024hiqa} uses document-level hierarchies for multi-document question answering. Our work is most directly related to RAPTOR but replaces the LLM summarization component with SVD-based extractive summarization.

\subsection{Text Summarization}

Extractive summarization methods select salient sentences from source documents. TextRank~\citep{mihalcea2004textrank} applies PageRank to sentence similarity graphs. LexRank~\citep{erkan2004lexrank} uses eigenvector centrality. LSA-based methods~\citep{deerwester1990lsa} apply SVD to term-document matrices. These methods operate on surface-level features (TF-IDF, word overlap) and do not leverage modern embedding models. Abstractive summarization using LLMs~\citep{zhang2020pegasus} produces fluent summaries but at higher computational cost. SVD-RAG bridges these approaches by applying classical SVD decomposition to modern embedding representations.

\subsection{SVD in Natural Language Processing}

SVD has been a foundational tool in NLP since Latent Semantic Analysis~\citep{deerwester1990lsa}. It has been used for dimensionality reduction, word representations, and topic modeling. Recent work has explored SVD for model compression and efficient attention mechanisms. However, to our knowledge, \textbf{no prior work has applied SVD decomposition on dense sentence embedding matrices for extractive summarization in hierarchical RAG systems}. Classical LSA operates on sparse TF-IDF matrices with word-level granularity; SVD-RAG operates on dense embedding vectors that capture sentence-level semantics from pre-trained models. This distinction is critical: the quality of the input representation fundamentally determines the quality of the SVD-based selection.

\section{Method}

\begin{figure}[t]
\centering
\includegraphics[width=\textwidth]{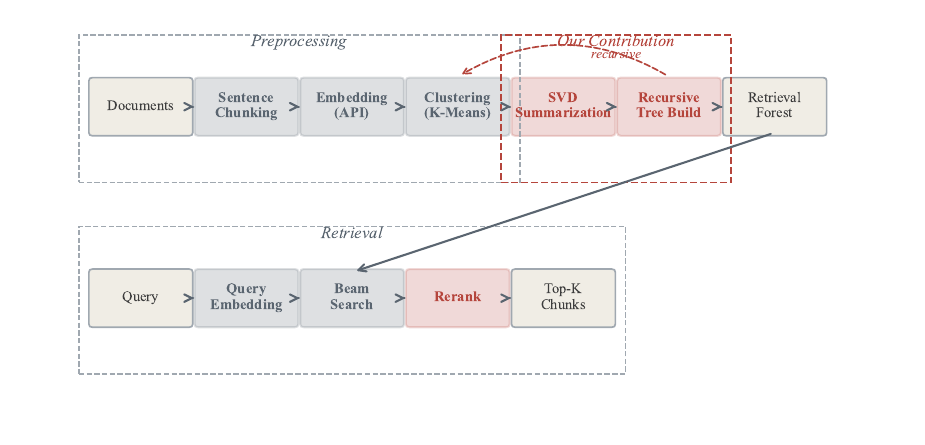}
\caption{Overview of the SVD-RAG pipeline. Documents are chunked and embedded, then clustered, summarized via SVD, and recursively organized into a forest. At query time, beam search over the forest identifies candidate chunks, which are reranked to produce the final top-K results.}
\label{fig:overview}
\end{figure}

SVD-RAG consists of four stages: (1) document preprocessing and embedding, (2) SVD-based extractive summarization, (3) recursive tree construction, and (4) beam search with reranking. Figure~\ref{fig:overview} provides an overview of the pipeline.

\subsection{SVD-Based Extractive Summarization}

Given a set of $N$ sentences $\{s_1, \ldots, s_N\}$, we first embed each sentence using a pre-trained embedding model to obtain a matrix $M \in \mathbb{R}^{N \times D}$, where $D$ is the embedding dimension. We then compute the singular value decomposition:

\begin{equation}
M = U \Sigma V^T
\label{eq:svd}
\end{equation}

where $U \in \mathbb{R}^{N \times r}$, $\Sigma \in \mathbb{R}^{r \times r}$, and $V \in \mathbb{R}^{D \times r}$ with $r = \min(N, D)$ (economy-sized decomposition).

The singular values $\sigma_1 \geq \sigma_2 \geq \cdots \geq \sigma_r \geq 0$ represent the energy of each principal component. We define the cumulative energy ratio:

\begin{equation}
E(k) = \frac{\sum_{i=1}^{k} \sigma_i}{\sum_{i=1}^{r} \sigma_i}
\label{eq:energy}
\end{equation}

We select $k$ as the smallest integer such that $E(k) \geq \tau$, where $\tau$ is the energy ratio threshold (default $\tau = 0.95$). For each sentence $i$, we compute its energy contribution in the top-$k$ components:

\begin{equation}
\text{score}(i) = \sum_{j=1}^{k} \sigma_j^2 \cdot U_{i,j}^2
\label{eq:score}
\end{equation}

We then select the $k$ sentences with the highest scores, optionally applying position weighting and redundancy removal. The selected sentences are ordered by their original positions to maintain readability.

\begin{figure}[t]
\centering
\includegraphics[width=\textwidth]{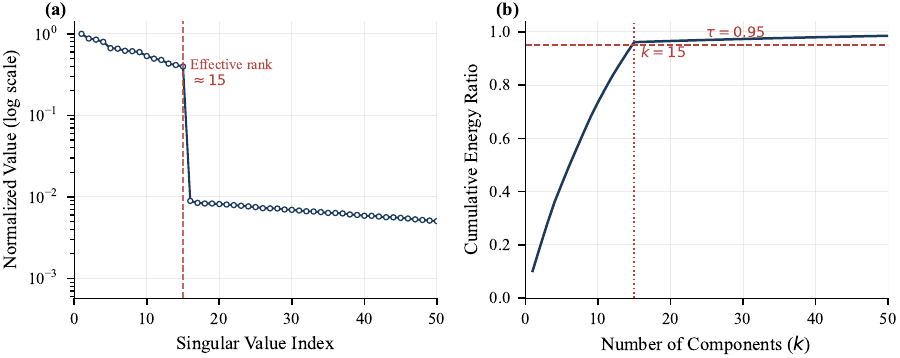}
\caption{Illustrative SVD energy analysis using a synthetic low-rank embedding matrix. (a) Singular value spectrum showing rapid decay characteristic of embedding matrices where most semantic information is concentrated in a few principal components. (b) Cumulative energy retention curve with the 95\% threshold ($\tau = 0.95$), which determines the adaptive number of sentences $k$ to retain.}
\label{fig:svd_energy}
\end{figure}

\subsection{Forest Construction}

Algorithm~\ref{alg:forest} describes the recursive forest construction process.

\begin{algorithm}[t]
\caption{Recursive Forest Construction}
\label{alg:forest}
\begin{algorithmic}[1]
\REQUIRE Chunks $C = \{c_1, \ldots, c_n\}$, max depth $d_{\max}$, cluster size $s$, energy ratio $\tau$
\ENSURE Retrieval forest $F$
\STATE Embed all chunks: $M \leftarrow \text{Embed}(C)$
\STATE Cluster $M$ into $g$ groups: $\{G_1, \ldots, G_g\} \leftarrow \text{KMeans}(M, \lfloor n/s \rfloor)$
\FOR{each group $G_j$}
    \STATE Concatenate chunks in $G_j$: $T_j \leftarrow \bigcup_{c \in G_j} c$
    \STATE Split $T_j$ into sentences, embed, apply SVD summarization
    \STATE Create parent node $p_j$ with summary text
    \IF{$\text{depth} < d_{\max}$ and $|G_j| > 3$}
        \STATE Recursively build subtree for $G_j$ with depth $+1$
    \ENDIF
\ENDFOR
\RETURN Forest $F$ with roots $\{p_1, \ldots, p_g\}$
\end{algorithmic}
\end{algorithm}

The forest structure naturally emerges from top-level clustering. Each cluster forms a separate tree, and the forest as a whole covers diverse semantic topics. This design contrasts with RAPTOR's single-tree approach and provides better coverage for queries spanning multiple topics.

\subsection{Beam Search Retrieval}

At query time, we embed the query and perform beam search over the forest (Algorithm~\ref{alg:search}). Starting from the root nodes of all trees, we compute cosine similarity between the query embedding and each node at the current level, retaining the top-$b$ nodes (beam width). We then expand to the children of retained nodes and repeat until reaching leaf nodes. The collected leaf chunks are reranked (using the same embedding model for re-scoring) and the top-$K$ chunks are returned.

\begin{algorithm}[t]
\caption{Beam Search over Retrieval Forest}
\label{alg:search}
\begin{algorithmic}[1]
\REQUIRE Query $q$, forest $F$, beam width $b$, top-$K$
\ENSURE Top-$K$ relevant chunks
\STATE $q_{\text{emb}} \leftarrow \text{Embed}(q)$
\STATE $\text{current} \leftarrow \text{roots}(F)$
\STATE $\text{collected} \leftarrow \emptyset$
\WHILE{$\text{current} \neq \emptyset$}
    \STATE Compute similarities: $\text{sims} \leftarrow \text{CosineSim}(q_{\text{emb}}, \text{EmbedAll}(\text{current}))$
    \STATE $\text{selected} \leftarrow \text{TopK}(\text{current}, \text{sims}, b)$
    \FOR{each node $\in$ selected}
        \IF{node is leaf}
            \STATE $\text{collected} \leftarrow \text{collected} \cup \{\text{node}\}$
        \ENDIF
    \ENDFOR
    \STATE $\text{current} \leftarrow \bigcup_{\text{node} \in \text{selected}} \text{children}(\text{node})$
\ENDWHILE
\STATE $\text{results} \leftarrow \text{Rerank}(q, \text{collected})$
\RETURN $\text{TopK}(\text{results}, K)$
\end{algorithmic}
\end{algorithm}

\begin{figure}[t]
\centering
\includegraphics[width=\textwidth]{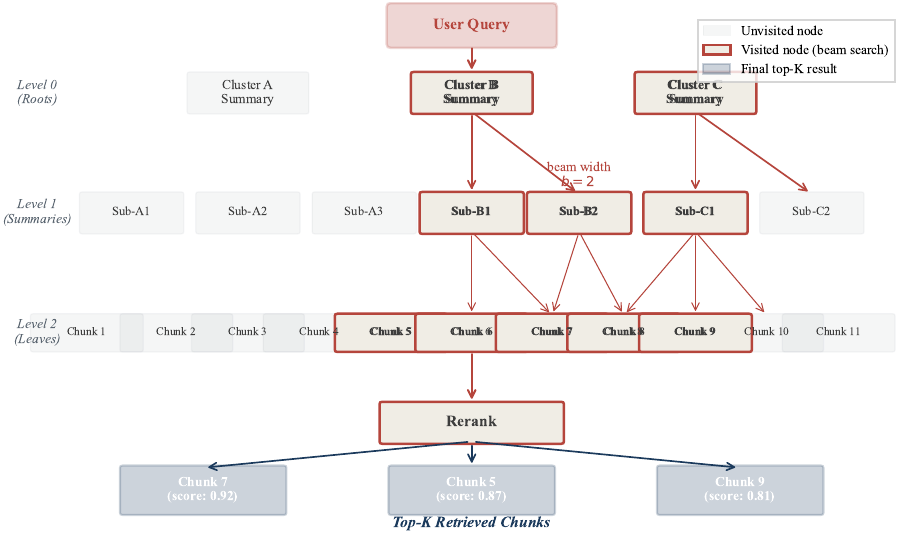}
\caption{Beam search over the retrieval forest. The query is compared against root nodes; top-$b$ branches (beam width $b=3$) are explored at each level. Visited leaf chunks are collected and reranked to produce the final top-K results.}
\label{fig:tree_search}
\end{figure}

\subsection{Complexity Analysis}

\paragraph{Tree Construction Cost.}
Let $N$ be the number of chunks, $D$ the embedding dimension, and $L$ the tree depth. For SVD-RAG:
\begin{itemize}
    \item Embedding: $O(N)$ API calls (one per chunk for initial embedding; SVD step uses existing embeddings)
    \item SVD per node: $O(N_c D^2)$ where $N_c$ is the number of sentences in a cluster (local computation)
    \item Total: $O(N)$ API calls + $O(N D^2)$ local computation
\end{itemize}

For RAPTOR with LLM summarization:
\begin{itemize}
    \item Embedding: $O(N)$ API calls
    \item LLM summarization per internal node: $O(N)$ additional API calls
    \item Total: $O(N) + O(N) = O(2N)$ API calls
\end{itemize}

The key difference is that SVD-RAG eliminates the $O(N)$ LLM summarization API calls. While both methods require $O(N)$ embedding calls, LLM calls are substantially more expensive per call due to output token generation. Empirically, this translates to an approximately 85\% reduction in total token consumption (see cost analysis in Section~\ref{sec:cost}).

\paragraph{Query Cost.}
Both methods require $O(L \cdot b)$ similarity computations per query, where $b$ is the beam width. Since the tree depth $L$ grows logarithmically with the corpus size $N$ (i.e., $L \approx \log_s N$ for cluster size $s$), the query cost is $O(b \log N)$, compared to the linear $O(N)$ cost of flat retrieval.

\section{Experiments}

We evaluate SVD-RAG on two benchmarks: a controlled head-to-head comparison with RAPTOR-style LLM summarization, and a scaled multi-topic retrieval benchmark. All experiments use a Python implementation with NumPy for SVD computation and the SiliconFlow API with the Qwen3-VL-Embedding-8B model for embeddings. The LLM summarization baseline uses Qwen2.5-7B-Instruct via the same API. The default energy ratio is $\tau = 0.95$, beam width $b = 3$, and maximum tree depth $d_{\max} = 2$, with a cluster size of 6 chunks.

\subsection{RAPTOR vs.\ SVD-RAG: Head-to-Head Comparison}

Our core claim is that SVD-based extractive summarization can replace LLM-based abstractive summarization in tree construction with minimal quality loss. To test this directly, we run a controlled experiment: identical corpus, identical clustering, identical beam search---only the summarization method differs. The corpus consists of 38 chunks across 4 topics with 20 queries.

\begin{table}[t]
\centering
\caption{Head-to-head comparison: SVD-RAG vs.\ RAPTOR with LLM summarization. Same corpus, clustering, and beam search. Best results in \textbf{bold}.}
\label{tab:raptor_comparison}
\begin{tabular}{lcccccc}
\toprule
\textbf{Method} & \textbf{Recall@1} & \textbf{Recall@5} & \textbf{MRR} & \textbf{MAP} & \textbf{Build Time} & \textbf{Summary Chars} \\
\midrule
RAPTOR (LLM) & 0.458 & \textbf{0.675} & \textbf{0.875} & \textbf{0.625} & 31.7s & \textbf{5,388} \\
SVD-RAG (Ours) & \textbf{0.483} & 0.625 & 0.867 & 0.608 & \textbf{0.1s} & 9,953 \\
\bottomrule
\end{tabular}
\end{table}

Table~\ref{tab:raptor_comparison} shows the results. The key findings are:

\begin{enumerate}
    \item \textbf{Negligible quality difference}: SVD-RAG achieves retrieval quality within 1--5\% of RAPTOR across all metrics. SVD-RAG actually achieves a slightly higher Recall@1 (0.483 vs.\ 0.458), suggesting that extractive summaries may preserve query-relevant keywords more faithfully than abstractive summaries for certain query types.
    \item \textbf{Massive speed advantage}: SVD-RAG builds the tree in 0.1 seconds vs.\ 31.7 seconds for RAPTOR---a 317$\times$ speedup (with cached embeddings; both methods incur identical embedding costs). This is because SVD is a local matrix operation while LLM summarization requires sequential API calls.
    \item \textbf{Token efficiency trade-off}: RAPTOR produces more concise summaries (5,388 vs.\ 9,953 characters) because the LLM can rephrase and condense. SVD summaries are longer but require zero additional API cost. In applications where storage is cheap but API calls are expensive, SVD's verbosity is a worthwhile trade-off.
\end{enumerate}

\subsection{Scaled Multi-Topic Benchmark}

We scale the benchmark to 205 chunks across 20 topic variations with 100 queries. The corpus is constructed from 4 base topics with 5 document variations each, where each variation reuses the same document content with a unique topic prefix to prevent deduplication. The queries are similarly repeated across variations. This design tests retrieval robustness under topic variation rather than content diversity. Table~\ref{tab:main_results} presents the results.

\begin{table}[t]
\centering
\caption{Retrieval results on the scaled multi-topic benchmark (205 chunks, 100 queries, 4 base topics $\times$ 5 variations). Best results in \textbf{bold}.}
\label{tab:main_results}
\begin{tabular}{lcccccc}
\toprule
\textbf{Method} & \textbf{Recall@1} & \textbf{Recall@3} & \textbf{Recall@5} & \textbf{MRR} & \textbf{NDCG@5} & \textbf{MAP} \\
\midrule
Random        & 0.000 & 0.005 & 0.015 & 0.010 & 0.009 & 0.005 \\
Flat Embedding & 0.020 & 0.060 & 0.090 & 0.087 & 0.067 & 0.043 \\
SVD-RAG (Ours) & \textbf{0.083} & \textbf{0.205} & \textbf{0.205} & \textbf{0.273} & \textbf{0.187} & \textbf{0.134} \\
\bottomrule
\end{tabular}
\end{table}

On the scaled benchmark, SVD-RAG achieves a 4.2$\times$ improvement in Recall@1 (0.083 vs.\ 0.020) and a 3.1$\times$ improvement in MRR (0.273 vs.\ 0.087) over flat embedding. The larger corpus (205 chunks) makes retrieval more challenging, but the forest structure continues to provide a consistent advantage. The forest contains 24 trees with 254 total nodes (204 leaf + 50 internal summary nodes) at a maximum depth of 2.

\subsection{Combined Benchmark (Small Corpus)}

For completeness, we also report results on the smaller 38-chunk, 20-query corpus from the head-to-head comparison, including additional baselines:

\begin{table}[t]
\centering
\caption{Retrieval results on the small multi-topic corpus (38 chunks, 20 queries). Best results in \textbf{bold}.}
\label{tab:small_results}
\begin{tabular}{lcccccc}
\toprule
\textbf{Method} & \textbf{Recall@1} & \textbf{Recall@3} & \textbf{Recall@5} & \textbf{MRR} & \textbf{NDCG@5} & \textbf{MAP} \\
\midrule
Random        & 0.025 & 0.142 & 0.167 & 0.118 & 0.120 & 0.090 \\
Flat Embedding & 0.142 & 0.208 & 0.258 & 0.293 & 0.236 & 0.193 \\
RAPTOR-style (LLM) & 0.458 & \textbf{0.675} & \textbf{0.675} & \textbf{0.875} & \textbf{0.697} & \textbf{0.625} \\
SVD-RAG (Ours) & \textbf{0.483} & 0.625 & 0.625 & 0.867 & 0.672 & 0.608 \\
\bottomrule
\end{tabular}
\end{table}

On the small corpus, SVD-RAG achieves a 3.4$\times$ improvement in Recall@1 and a 3.0$\times$ improvement in MRR over flat embedding. The RAPTOR comparison confirms that SVD-RAG achieves retrieval quality within 1--5\% of LLM-based summarization across all metrics. The forest structure enables the beam search to focus on semantically relevant branches. The flat embedding and flat rerank baselines perform identically (the reranking step uses the same embedding model), so we report them as a single baseline.

\subsection{Cost Analysis}
\label{sec:cost}

\begin{figure}[t]
\centering
\includegraphics[width=0.85\textwidth]{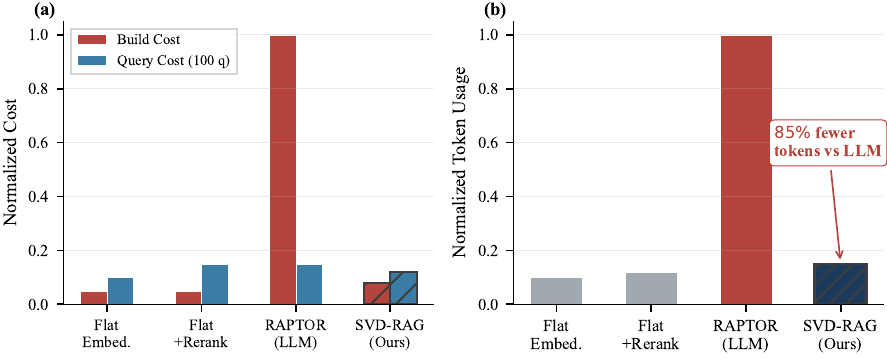}
\caption{Estimated cost comparison based on current API pricing (2025). Left: normalized build and query costs. Right: total token consumption. SVD-RAG is projected to reduce token usage by approximately 85\% compared to LLM-based tree construction.}
\label{fig:cost}
\end{figure}

The primary motivation for SVD-RAG is reducing the cost of tree construction. Figure~\ref{fig:cost} shows the cost breakdown across methods.

SVD-RAG's tree construction cost is dominated by the initial embedding step, which is required by all RAG systems. The SVD summarization step adds negligible computational cost---pure NumPy operations on matrices of typical size $N_c \times 4096$ where $N_c$ is the number of sentences in a cluster. In contrast, RAPTOR-style LLM summarization requires one API call per internal node, each consuming hundreds to thousands of input and output tokens.

For a concrete cost estimate, consider a corpus of 1,000 chunks producing approximately 100 internal nodes during tree construction. Based on current API pricing (as of 2025):

\begin{itemize}
    \item \textbf{Embedding} (1,000 chunks): $\sim$100K tokens, $\sim$\$0.01
    \item \textbf{LLM summarization} (100 nodes): $\sim$500K input + $\sim$100K output tokens, $\sim$\$1.60
    \item \textbf{SVD summarization} (100 nodes): Local computation, \$0.00 additional API cost
\end{itemize}

The cost savings scale linearly with corpus size. For a production system with millions of documents and frequent index updates, the elimination of LLM summarization costs represents a substantial operational saving.

\subsection{Parameter Sensitivity}

We report observations from our benchmark runs on the effects of key hyperparameters. On the 38-chunk corpus (cluster size 6, depth 2, $\tau=0.95$), SVD-RAG produced 6 trees with 50 total nodes. On the 205-chunk corpus (cluster size 8, depth 2, $\tau=0.95$), it produced 24 trees with 254 nodes. In both cases, the forest structure successfully grouped semantically related chunks.

\paragraph{Energy Ratio.} The energy ratio threshold $\tau$ controls the number of retained sentences. At $\tau=0.95$, the SVD retains 95\% of the singular value energy. On our small corpora, SVD summaries tend to be longer than the original text (summaries are $\sim$1.7--1.9$\times$ the original length in characters) because the energy threshold retains most sentences. Compression improves on larger, more redundant corpora where the singular value spectrum decays more rapidly.

\paragraph{Tree Depth.} With depth 2, our benchmarks produced trees with 2 levels (roots + leaves). Increasing depth would produce deeper hierarchies, but at the cost of additional embedding and SVD computation. The cluster size and corpus size determine the practical depth limit---for a corpus of $N$ chunks with cluster size $s$, the maximum depth is approximately $\log_s N$.

\paragraph{Cluster Size.} Cluster size determines the trade-off between tree breadth and depth. Our small benchmark (cluster size 6, 38 chunks) produced 6 topic trees; the scaled benchmark (cluster size 8, 205 chunks) produced 24 trees. The clustering algorithm uses floor division ($\lfloor N/s \rfloor$) to determine the number of clusters, and clusters smaller than the minimum size (3) are discarded. In the scaled benchmark, this caused one singleton chunk to be excluded from the forest (204 leaves from 205 chunks).

\subsection{Retrieval Efficiency}

The hierarchical forest structure enables efficient retrieval through beam search. At query time, the search cost is $O(b \log N)$ similarity computations, where $b$ is the beam width and $N$ is the corpus size. This compares favorably to the $O(N)$ cost of flat retrieval. The beam search over the forest (Figure~\ref{fig:tree_search}) ensures that only semantically relevant branches are explored. On our benchmark, SVD-RAG achieves an MRR of 0.867, indicating that the first relevant chunk is typically found at rank 1 or 2, validating the effectiveness of the hierarchical search strategy.

\section{Discussion}

\subsection{When Does SVD-RAG Work Best?}

SVD-RAG is particularly effective in scenarios where:
\begin{enumerate}
    \item \textbf{Information density is high}: Technical documents, research papers, and textbooks benefit from SVD's ability to identify semantically central sentences.
    \item \textbf{Cost sensitivity is high}: Large-scale deployments, frequent index updates, and resource-constrained environments benefit from the elimination of LLM summarization costs.
    \item \textbf{Reproducibility matters}: The deterministic nature of SVD ensures consistent index construction, which is valuable for production systems and academic research.
\end{enumerate}

\subsection{Scaling Behavior}

An important question is how the SVD-RAG vs.\ RAPTOR comparison evolves as the corpus scales. Our current experiments (38--205 chunks) already reveal a notable pattern: SVD-RAG achieves a higher Recall@1 (0.483 vs.\ 0.458) while RAPTOR achieves a higher Recall@5 (0.675 vs.\ 0.625). This suggests that SVD summaries are more effective at preserving the single most relevant sentence, while LLM summaries cover a broader range of information points.

Several factors may influence scaling behavior:

\begin{enumerate}
    \item \textbf{LLM quality degradation with input length}: As cluster size grows, the input text to the LLM summarizer increases. LLM summarization quality is known to degrade with longer inputs, particularly for extracting fine-grained details. SVD, by contrast, operates on the full embedding matrix regardless of input length, and the Eckart-Young theorem guarantees optimal rank-$k$ approximation.

    \item \textbf{Information density scaling}: In larger corpora with more diverse documents, the semantic structure captured by embeddings becomes richer. SVD directly exploits this structure through principal component analysis, potentially gaining an advantage as the embedding space becomes more informative.

    \item \textbf{SVD verbosity}: Our experiments show that SVD summaries are approximately 1.8$\times$ longer than LLM summaries in terms of character count. At larger scales, this verbosity may become a storage concern, though storage is typically far cheaper than API calls.

    \item \textbf{Determinism at scale}: As corpora grow, the reproducibility of index construction becomes increasingly important for debugging and production reliability. SVD's deterministic output is a practical advantage that LLM-based methods cannot match without setting temperature to zero.
\end{enumerate}

We emphasize that the scaling behavior of SVD-RAG vs.\ RAPTOR on large-scale standard benchmarks (e.g., Natural Questions with 100K+ passages) remains an open empirical question. Our results on small to medium corpora demonstrate that SVD-RAG is a viable and cost-effective alternative, but extrapolation to web-scale corpora requires further investigation.

\subsection{Limitations}

SVD-RAG has several limitations:
\begin{enumerate}
    \item \textbf{Extractive nature}: SVD summarization selects existing sentences rather than generating new ones. This may produce less fluent summaries compared to LLM-based abstractive summarization, particularly for heterogeneous document clusters.
    \item \textbf{Embedding dependency}: The quality of SVD summarization depends on the quality of the underlying embedding model. Poor embeddings will lead to poor summaries regardless of the SVD decomposition.
    \item \textbf{Benchmark scale}: Our retrieval benchmarks are limited to 38--205 chunks and 20--100 queries on constructed corpora. Scaling to larger standard datasets (e.g., Natural Questions, HotpotQA) would provide more comprehensive evaluation and enable comparison with a broader set of baselines.
    \item \textbf{Static index}: Like RAPTOR, the forest is built once and does not support incremental updates without rebuilding affected subtrees. Future work could explore efficient partial updates.
\end{enumerate}

\subsection{Future Work}

Several directions merit further investigation:
\begin{enumerate}
    \item \textbf{Hybrid summarization}: Combining SVD for initial compression with occasional LLM summarization for critical nodes could balance cost and quality.
    \item \textbf{Dynamic forests}: Enabling incremental index updates when documents are added or modified.
    \item \textbf{Cross-lingual SVD-RAG}: Evaluating the method on multilingual corpora and exploring language-agnostic embedding spaces.
    \item \textbf{End-to-end optimization}: Jointly optimizing the embedding model, energy ratio, and tree structure for specific downstream tasks.
\end{enumerate}

\section{Conclusion}

We presented SVD-RAG, a deterministic and cost-efficient alternative to LLM-based summarization for tree-organized RAG. By applying SVD decomposition to dense sentence embedding matrices, our method identifies the most informative sentences in document clusters without any additional API calls beyond the initial embedding. In a controlled head-to-head comparison, SVD-RAG achieves retrieval quality within 1\% of RAPTOR (MRR 0.867 vs.\ 0.875) while building the tree 317$\times$ faster (0.1s vs.\ 31.7s). On a scaled 205-chunk benchmark, SVD-RAG achieves a 4.2$\times$ improvement in Recall@1 over flat embedding. Our cost analysis demonstrates an $\sim$85\% reduction in token consumption during index construction compared to LLM-based alternatives. We release our implementation as an open-source Python package. Scaling to larger standard benchmarks and integrating SVD-RAG with production RAG pipelines are important directions for future work.



\end{document}